\documentclass[twocolumn]{aastex631}

\submitjournal{AJ}

\shorttitle{The IMFs of seven YMCs in NGC 4038/9}
\shortauthors{Koo et al.}

\usepackage{ulem}

\begin{document}

\title{Initial Mass Functions of Young Stellar Clusters from the Gemini Spectroscopic Survey of Nearby Galaxies I. Young Massive Clusters in the Antennae galaxies}

\correspondingauthor{Beomdu Lim}
\email{blim@kongju.ac.kr}

\author[0000-0001-8969-0009]{Jae-Rim Koo}
\affiliation{Earth Environment Research Center, Kongju National University, 56 Gongjudaehak-ro, Gongju-si, Chungcheongnam-do 32588, Republic of Korea}

\author[0000-0001-9263-3275]{Hyun-Jeong Kim}
\affiliation{Earth Environment Research Center, Kongju National University, 56 Gongjudaehak-ro, Gongju-si, Chungcheongnam-do 32588, Republic of Korea}
\affiliation{Korea Astronomy and Space Science Institute, 776 
Daedeok-daero, Yuseong-gu, Daejeon 34055, Republic of Korea}

\author[0000-0001-5797-9828]{Beomdu Lim}
\affiliation{Earth Environment Research Center, Kongju National University, 56 Gongjudaehak-ro, Gongju-si, Chungcheongnam-do 32588, Republic of Korea}
\affiliation{Korea Astronomy and Space Science Institute, 776 
Daedeok-daero, Yuseong-gu, Daejeon 34055, Republic of Korea}
\affiliation{Department of Earth Science Education, Kongju National University, 
56 Gongjudaehak-ro, Gongju-si, Chungcheongnam-do 32588, Republic of Korea}



\begin{abstract} 
The stellar initial mass function (IMF) is a key parameter to understand 
the star formation process and the integrated properties of stellar 
populations in remote galaxies. We present a spectroscopic study 
of young massive clusters (YMCs) in the starburst galaxies NGC 4038/39. 
The integrated spectra of seven YMCs obtained with GMOS-S attached to 
the 8.2-m Gemini South telescope reveal the spectral features associated 
with stellar ages and the underlying IMFs. We constrain the ages of 
the YMCs using the absorption lines and strong emission bands 
from Wolf-Rayet stars. The internal reddening is also estimated from the 
strength of the \ion{Na}{1} D absorption lines. Based on these constraints, 
the observed spectra are matched with the synthetic spectra generated 
from a simple stellar population model. Several parameters of the 
clusters including age, reddening, cluster mass, and the underlying IMF 
are derived from the spectral matching. The ages of the YMCs range 
from 2.5 to 6.5 Myr, and these clusters contain stellar masses 
ranging from $1.6\times10^5 \ M_{\sun}$ to $7.9\times10^7 \ M_{\sun}$. 
The underlying IMFs appear to differ from the universal form of 
the Salpeter/Kroupa IMF. Interestingly, massive clusters 
tend to have the bottom-heavy IMFs, although the masses of some clusters 
are overestimated due to the crowding effect. Based on this, our results 
suggest that the universal form of the IMF is not always 
valid when analyzing integrated light from unresolved stellar 
systems. However, further study with a larger sample size 
is required to reach a definite conclusion.
\end{abstract}



\section{Introduction} \label{sec:1}
Stellar systems serve as laboratories to study the star formation 
process, because approximately 80--90 \% stars form in stellar clusters 
or associations \citep{LL03,PCA03} in the Galaxy. In particular, 
young massive clusters (YMCs) are the nests of massive stars that are 
very rare in the Solar neighborhood. The conditions of massive star 
formation can be studied statistically using the initial 
mass function (IMF) sampled from a wide range of stellar masses 
\citep{LSH17}. In addition, stellar systems generally form 
along the spiral arms of host galaxies and are dissolved into galactic 
disks \citep{BSD01,LL03,G18}. Therefore, they play a crucial role in determining 
the properties of stellar population in their host galaxies.

A large number of observational studies have been performed on 
stellar clusters in the Galaxy, because the individual members can be 
spatially resolved due to their proximity. However, the severely 
uneven distribution of the interstellar medium across the Galactic 
disk introduces observational bias in the cluster sample. 
Several YMCs in the inner Galaxy have been steadily 
studied in the infrared domain \citep[etc.]{FKM99,FMR06,DFK07,BCS08,CND09,LCS13,HLA19}, 
but most cluster samples are limited to within 3 kpc from the 
Sun \citep{CJV18,DMM21}. This observational bias limits our ability 
to study star formation in different environments.

Stellar clusters in nearby galaxies can provide constraints on 
the formation process of stars and clusters when combined with 
the Galactic sample. Such stellar 
systems are generally found in molecular gas disks 
\citep{TBM07,HL08,BTK09,WCS10,LHL13}. Interestingly, a large number 
of massive clusters tend to be located in extreme environments 
of interacting galaxies. For instance, \citet{dAB03} 
discovered high-mass clusters in the ring of NGC 3310 and the tidal 
tail of NGC 6745. \citet{TBS07} performed spectroscopic observations of three 
clusters found in the tidal tail of NGC 3256, which revealed 
the masses as high as $\sim 10^5 \ M_{\sun}$ for these clusters. 
\citet{BHK05} discovered many YMCs in the tidal tails of NGC 6872. 
These observational results suggest that such extreme environments 
are likely the preferential sites of massive cluster formation.

NGC 4038 and 4039 (NGC 4038/9), the so-called Antennae galaxies, are 
interacting galaxies that provide an excellent testbed 
for studying the star formation process in extreme environments 
subject to external forces. These galaxies 
contain a large amount of molecular gas 
\citep[$\sim 10^8 - 10^{10} \ M_{\sun}$;][]{SSS90,GLL01,
WSM03,ZSK03,SHM07}. A large fraction of H {\scriptsize I} 
gas ($\sim$ 70\%) is distributed in the tidal tails 
\citep{vdH79,HvdHB01}. The gas kinematics of these galaxies 
were investigated in several studies \citep{BB66,AMB92,HvdHB01,
SHM07}. \citet{NU00} detected a number of radio sources. 
About 30\% of the strongest radio sources are likely 
associated with active star-forming regions. Later, radio 
observations with the Atacama Large Millimeter/submillimeter 
Array newly identified some molecular structures and compact 
sources \citep{EKM12,WBC14}. 

A deep Chandra X-ray image revealed hot diffuse gas across 
NGC 4038/9, showing various levels of chemical 
enrichment \citep{FKZ03,FBK04}. A fraction of hot gas 
appears to be heated by feedback 
from massive stars in stellar clusters \citep{MCG04}. 
\citet{ZFB06} detected several tens of X-ray sources 
from an extensive monitoring survey, finding a flat X-ray 
luminosity function in their later study \citep{ZFB07}. Thus, most results from 
previous studies confirm that star formation is actively 
taking place in NGC 4038/9.

\citet{WS95} discovered more than 700 cluster candidates 
from Hubble Space Telescope (HST) observations, the brightest of which were considered 
young globular clusters. They also studied the clusters in 
NGC 4038/9 using higher resolution images taken with the Advanced 
Camera for Surveys and Near Infrared Camera and Multi-Object 
Spectrometer on board the HST \citep{WCS10}. Their infrared 
images revealed that approximately 16\% of the clusters are still embedded 
in their natal clouds. The spatial age distributions of the stellar 
clusters show signs of triggered cluster formation by older 
clusters. \citet{BTK09} performed multi-object spectroscopy of 
16 stellar clusters in the same galaxies with supplemental HST 
imaging in the optical domain. They postulated that most clusters 
formed after the merger given the range of their ages (3 -- 200 Myr). 

NGC 4038/9 are among the closest pairs of interacting galaxies, 
but not close enough to resolve the stellar clusters into 
individual members. The published distance to these galaxies ranges 
from 20.0 Mpc to 28.8 Mpc \citep{WS95,SBM08} depending 
on the adopted Hubble constant. Therefore, previous 
photometric and spectroscopic studies on stellar clusters 
investigated their physical properties based on the integrated 
light. In order to interpret the integrated light, it 
is essential to assume the IMF, because there is a degeneracy 
among underlying physical parameters. 

Several studies have shown that stellar IMFs, in general, do 
not deviate significantly from the Salpeter/Kroupa IMFs \citep{BCM10}, 
but there are signs of IMF variations \citep[and 
references therein]{LSH17,S20,B23,CKW24}. We initiated a 
systematic survey of stellar clusters in nearby galaxies 
to test the universality of stellar IMFs. NGC 4038/9 are 
the first targets from which we infer the underlying IMFs from the 
integrated spectra of YMCs formed in 
starburst environments within the interacting galaxies.
The data used are described in Section~\ref{sec:2}. 
We present the results of this study in Section~\ref{sec:3} and 
discuss several sources  that may have influenced our analysis 
in Section~\ref{sec:4}. Finally, our results are summarized 
in Section~\ref{sec:5}.

\begin{figure}[t]
\epsscale{1.0}
\plotone{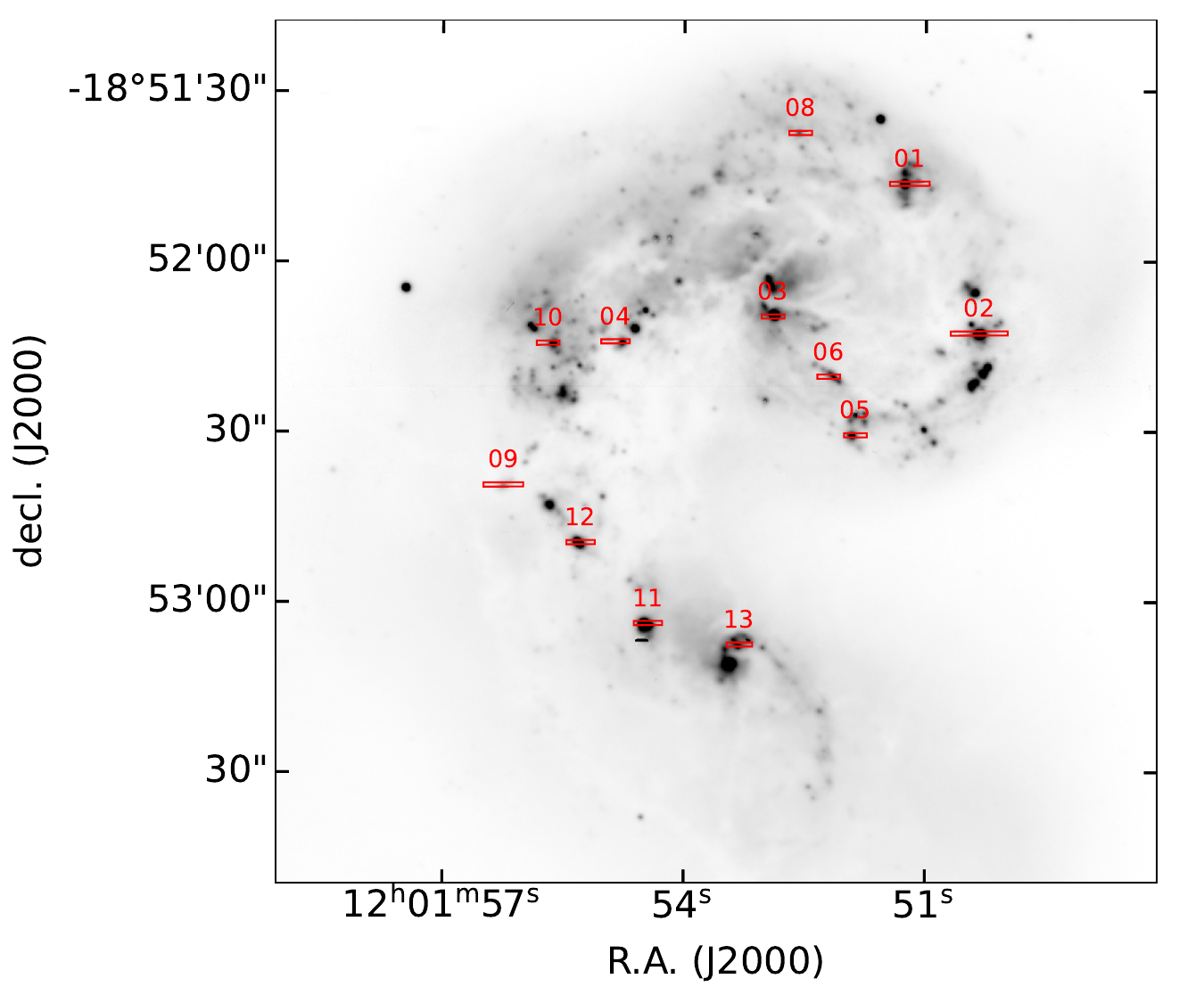}
\caption{GMOS-S $g^{\prime}$ band image of NGC 4038/9. The slit positions 
are shown by rectangular boxes with the cluster IDs.}\label{fig1}
\end{figure}

\section{Data} \label{sec:2}
We took the catalogs of 50 luminous clusters and 50 massive 
clusters from \citet{WCS10}. These two catalogs were merged 
into a single catalog after removing duplicates. A total 
of 29 clusters younger than 10 Myr and brighter than 
21.5 mag in the $V$ band were selected as candidates for 
spectroscopic observations. Indeed, these candidates 
have blue $U-B$ colors because they contain hot massive 
stars. Their total stellar masses range from $2\times10^4 
\ M_{\sun}$ to $1.6\times 10^6 \ M_{\sun}$ \citep{WCS10}.

\begin{figure*} \centering
\includegraphics[width=\textwidth]{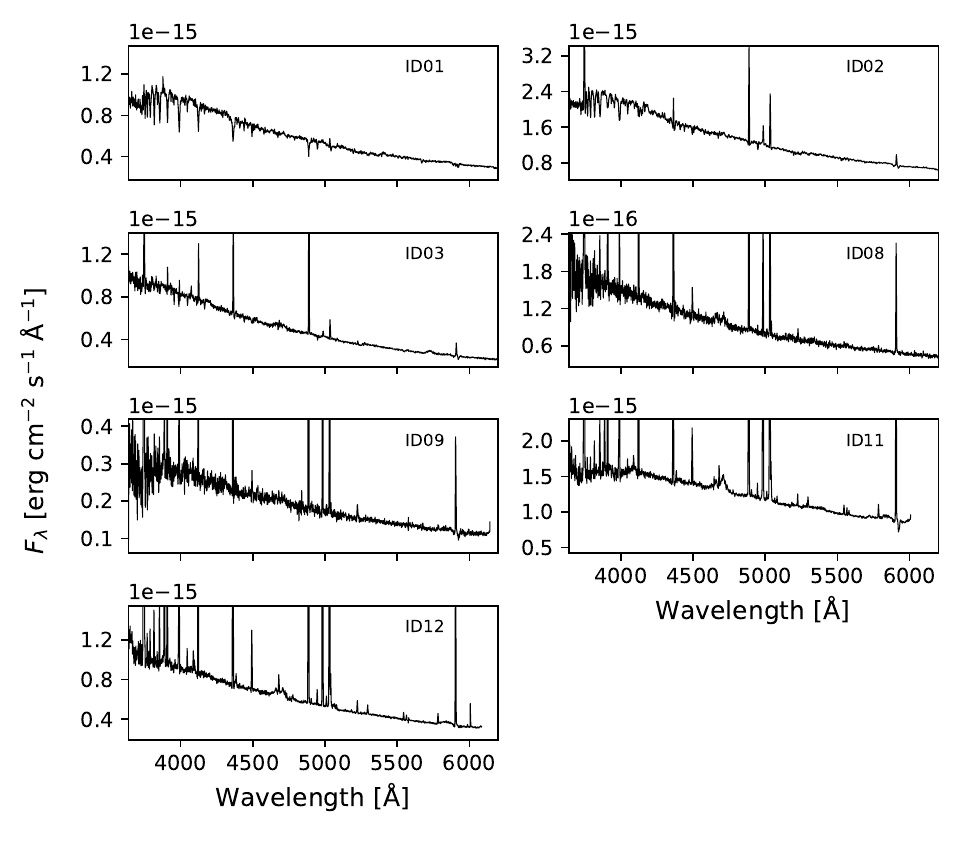}
\caption{Observed spectra of seven YMCs. The cluster ID 
is labeled in the upper right corner of each panel. The 
fluxes of the observed spectra were calibrated using the 
spectra of the standard star LTT 4816.}
\label{fig2}
\end{figure*}

\subsection{GMOS observation}\label{ssec:21}
We performed pre-imaging observations in the $g^{\prime}$ band 
and spectroscopic observations with the Gemini 
Multi-Object Spectrograph (GMOS)  
at the Gemini South observatory (GS-2022A-Q-122 and 
GS-2023A-Q-123). The observed images were used to design 
a mask for multi-object spectroscopy. We obtained the 
spectra of 12 YMCs out of 29 candidates to avoid overlapping 
spectra dispersed from adjacent slits in the mask. 
Figure~\ref{fig1} displays the positions of the observed YMCs 
on the GMOS pre-image. 

The B600 grating was used to cover a spectral range from 
3500 \AA \ to 6250 \AA, and the slit width was set to $0\farcs75$. 
In order to remove cosmic rays and to fill the two gaps 
of the GMOS detectors, we performed wavelength dithering by 
shifting the central wavelengths to 4200 \AA, 4500 \AA, and 
4800 \AA. A total of 35 GMOS frames were acquired over six 
nights from 2022 June 8 to 2023 February 16. The exposure time of 
each frame was 900 seconds. We note that the eight frames 
observed on 2022 June 8 were excluded 
from our analysis because the angular distance to the moon 
was closer than 30$\arcdeg$ (bright background) and the 
poor seeing condition. 
 
For flux calibration, the spectra of the standard 
star LTT 4816 were obtained using GMOS in a long-slit mode. 
The observations of the standard star were conducted on 
the same night as the target observing run on 2023 February 
13, with the same settings for the central wavelengths 
and slit-width.

All spectroscopic data were reduced using the Gemini 
IRAF package following the standard procedures. The 
solution for wavelength calibration was obtained 
from the arc spectra and applied to the spectra of 
the standard star and YMCs. The wavelength-corrected 
two-dimensional spectra were combined into a single 
spectrum for the same target observed on the same night. 
We extracted one-dimensional (1D) spectra using custom 
python scripts.  

In the slits, the observed YMCs have finite extents 
that are larger than the mean seeing of about 0\farcs7.
We fit a Gaussian profile to a spatial profile along the 
spatial axis of the slit at given wavelengths. Then, 
spectra were extracted within an aperture radius 
corresponding to 3$\sigma$ of the best-fit Gaussian 
profile. The median values of the signals from the outer 
part of each slit were considered as background 
signals. We subtracted these background signals 
from the extracted 1D spectra of the YMCs. 

The spatial profile of ID 13 is blended 
by the neighboring objects in the same slit. The 
slit length does not cover the full extent of 
the four YMCs (IDs 04, 05, 06, and 10) with a sufficiently 
wide sky background. Therefore, a total of five 
YMCs were excluded for further analysis. In this 
study, we analyzed the integrated spectra of seven 
YMCs. However, some spectra of the seven YMCs were 
additionally discarded because the slit length is still short compared 
to the observed spatial profile affected by the 
seeing variations among observing runs.

A small portion of the signals from the observed YMCs 
were lost due to the slit width being comparable to the 
diameter of the seeing disk.  In addition, atmospheric 
differential refraction occurs depending on the 
position angles of the slits on the sky. We corrected 
for these effects on the observed spectra using the 
\texttt{spec.lightloss2} function of 
\texttt{Ian's Astro-Python Code}\footnote{\url{https://crossfield.ku.edu/python/}}.

\begin{figure*} \centering
\includegraphics[width=\textwidth]{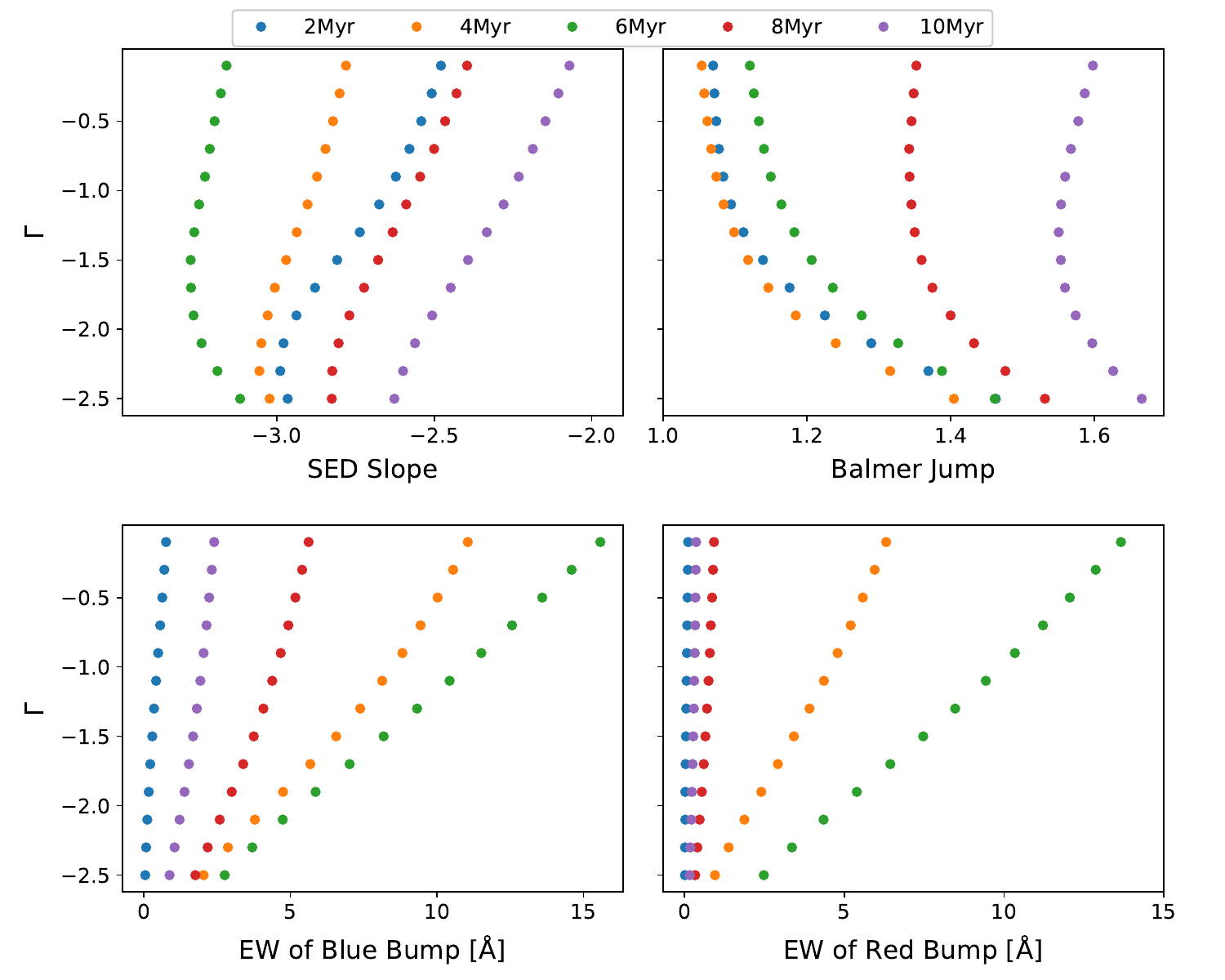}
\caption{Variations of spectral features obtained from the 
synthetic integrated spectra of model clusters with 
a total mass of $\log M_{\mathrm{cl}}$ = 6.0 and the Solar metallicity.
In the lower panels, the equivalent widths of blue and red bumps 
are obtained from the output of \texttt{STARBURST99}, 
\ion{N}{3} $\lambda$4640 + \ion{C}{3} $\lambda$4650 
+ \ion{He}{2} $\lambda$4686 and \ion{C}{4} $\lambda$5808, 
respectively.}
\label{fig3}
\end{figure*}

Recent quantum efficiency curves for the GMOS 
detector\footnote{The quantum efficiency curves 
of the GMOS detector adopted from Data Reduction for 
Astronomy from Gemini Observatory North and South 
(\texttt{DRAGONS}).} were used to correct for the 
quantum efficiency that varies over a wide range of 
wavelengths. For extinction correction, we applied 
the mean atmospheric extinction at the Gemini South 
observatory \citep[ctioextinction;][]{SB83}. The 
heliocentric radial velocity correction was also 
performed to combine the spectra observed on different 
nights. 

We used the spectra of the YMCs observed on the same 
night with the observations of the standard star LTT 4816 
as a reference. The spectra taken on different nights 
were scaled to the flux levels of the reference spectra 
and then combined into a single spectrum for individual cluster 
by adopting the median at given wavelengths. The final signal-to-noise 
ratios (SNRs) of the combined spectra range from 25 to 90 
at around 5400 \AA. Figure~\ref{fig2} displays the 
flux-calibrated spectra of the seven YMCs. 

To test the quality of our flux calibration, 
we conducted synthetic photometry for the flux-calibrated 
spectra of the standard star LTT 4816 and the four YMCs (IDs 01, 02, 
03 and 08) that are spatially well-isolated from bright sources using the 
\texttt{Pyphot} package \citep{F22}. The photometric 
zeropoint of $-0.2$ mag in the $V$ band was obtained by 
calculating the difference between the magnitudes from 
the SIMBAD database \citep{WOE00} and our synthetic 
photometry for the standard star. This zeropoint was then applied to the 
$V$ magnitudes of the four YMCs. We compared our 
photometric data with those of \citet{WCS10}. The 
mean difference between the two photometric data 
is less than 0.01 mag, and the standard deviation 
is about 0.32 mag. Therefore, we confirmed that the 
spectra of the YMCs are well calibrated in an absolute 
scale. 

\subsection{Simple stellar population model}\label{ssec:22}
It is essential to select a simple stellar population 
model to interpret the integrated light from unresolved 
stellar clusters in external galaxies. To this end, we adopted 
the synthetic model \texttt{STARBURST99} \citep{LSG99,LEM14} 
because of three reasons. First, it is possible to parameterize 
the underlying IMFs. This is the key option 
for the purpose of our survey. Second, \texttt{STARBURST99} 
adopts the latest evolutionary models for very massive 
stars up to 120 $M_{\sun}$ that take into account the 
effects of rotation on stellar evolution. Third, the synthetic 
model is constantly being improved, including new spectral 
libraries, new stellar atmospheric grids, supernova yields, 
etc.

In this study, we used the \texttt{GENEVA} stellar 
evolutionary models adopting a rotation rate 
$v_{\mathrm{ini}}/v_{\mathrm{break}}$ of 0.4 for the 
Solar ($Z = 0.014$) and sub-Solar ($Z = 0.002$) 
metallicities \citep{EGE12,GEE13}. Instantaneous 
star formation was considered as the star formation 
law for YMCs. The cluster masses ranged from 3 to 8 with 
an interval of 0.1 dex in the logarithmic scale ($\log M_{\mathrm{cl}}$). 
Stellar masses were generated 
in a range of 0.1 $M_{\sun}$ to 120 $M_{\sun}$. The 
frame of the underlying IMFs was adopted from the Kroupa 
IMF \citep{K01}, which is described by two power-law 
indices in two mass regimes: 0.1--0.5 $M_{\sun}$ and 
0.5--120 $M_{\sun}$. We varied the power-law index 
$\Gamma$ (or $1-\alpha$) for the higher mass regime 
from $0.0$ to $-2.5$ in 0.1 steps. Cluster ages were 
considered from 1 Myr to 15 Myr in 0.5 Myr intervals. 
The default settings were used for the other options. 
A total of 153,816 high resolution spectra were 
generated for all setups. These synthetic 
spectra cover a wide spectral range from 
3000 \AA \ to 7000 \AA. Finally, we degraded the 
spectral resolution of the synthetic spectra to 
that of the GMOS spectra ($R\sim1700$) using 
\texttt{SPECTRUM/SMOOTH2}\footnote{\url {https://www.appstate.edu/~grayro/spectrum/spectrum.html}} \citep{GC94}.

We analyzed the synthetic spectra generated 
from \texttt{STARBURST99} to find spectral 
features that depend on the age of clusters and 
the underlying IMFs. The continuum spectral 
energy distribution (SED) of an integrated 
spectrum may be related to the content of massive 
stars. We measured the SED slopes of all synthetic 
spectra in the wavelength range from 4100 to 5700 
\AA \ on a logarithmic scale using a least-square 
fitting method. The upper left panel of Figure~\ref{fig3} 
displays the relation between the SED slope and 
the power-law index $\Gamma$. The SED slope shows 
a smooth variation with the underlying IMFs at a given 
cluster age. 

The Balmer jump is a spectral feature related 
to the cluster age and underlying IMF. The integrated 
spectra of older clusters may have stronger Balmer 
jumps than those of younger clusters, because they contain more 
later-type stars. The strength of the Balmer jump 
is also related to the number ratios of O-type stars 
and later-type stars indicating the IMFs. We measured 
the strength of the Balmer jump from the synthetic 
spectra. In this study, the ratio of the pseudo-continuum 
flux and the actual flux between 3650 and 3690~\AA \ 
was defined as the strength of the Balmer jump. The 
upper right panel of Figure~\ref{fig3} shows the trend 
between the Balmer jump and power-law index $\Gamma$ as 
expected.

The YMCs younger than 8 Myr contain Wolf-Rayet stars. These 
massive evolved stars show characteristic spectral features 
originating from their winds in the optical passbands 
\citep{GC09}. The blue bump is the broad emission band 
composed of the blended \ion{N}{3} $\lambda4634$--41 
and \ion{He}{2} $\lambda$4686 emission lines, and the 
red bump is characterized by \ion{C}{4} $\lambda$5808. 
We obtained the equivalent widths (EWs) of the blue and 
red bumps from the output of \texttt{STARBURST99} for different cluster 
ages and underlying IMFs (the lower panels of 
Figure~\ref{fig3}). The blue bump is measurable 
for the first 8 Myr, while the red bump attenuates 
after 6 Myr. The two bumps appear to be getting stronger as 
the IMF becomes more top-heavy type ($\Gamma > 
- 1.3$).

These spectral features correlate with the underlying IMFs 
for a given age, but there is a degeneracy between the ages 
and the underlying IMFs for the same measure of given 
spectral features. 
Constraining cluster ages independently may mitigate this degeneracy.

\section{Results} \label{sec:3}
\begin{figure*}[t]
\epsscale{1.0}
\plotone{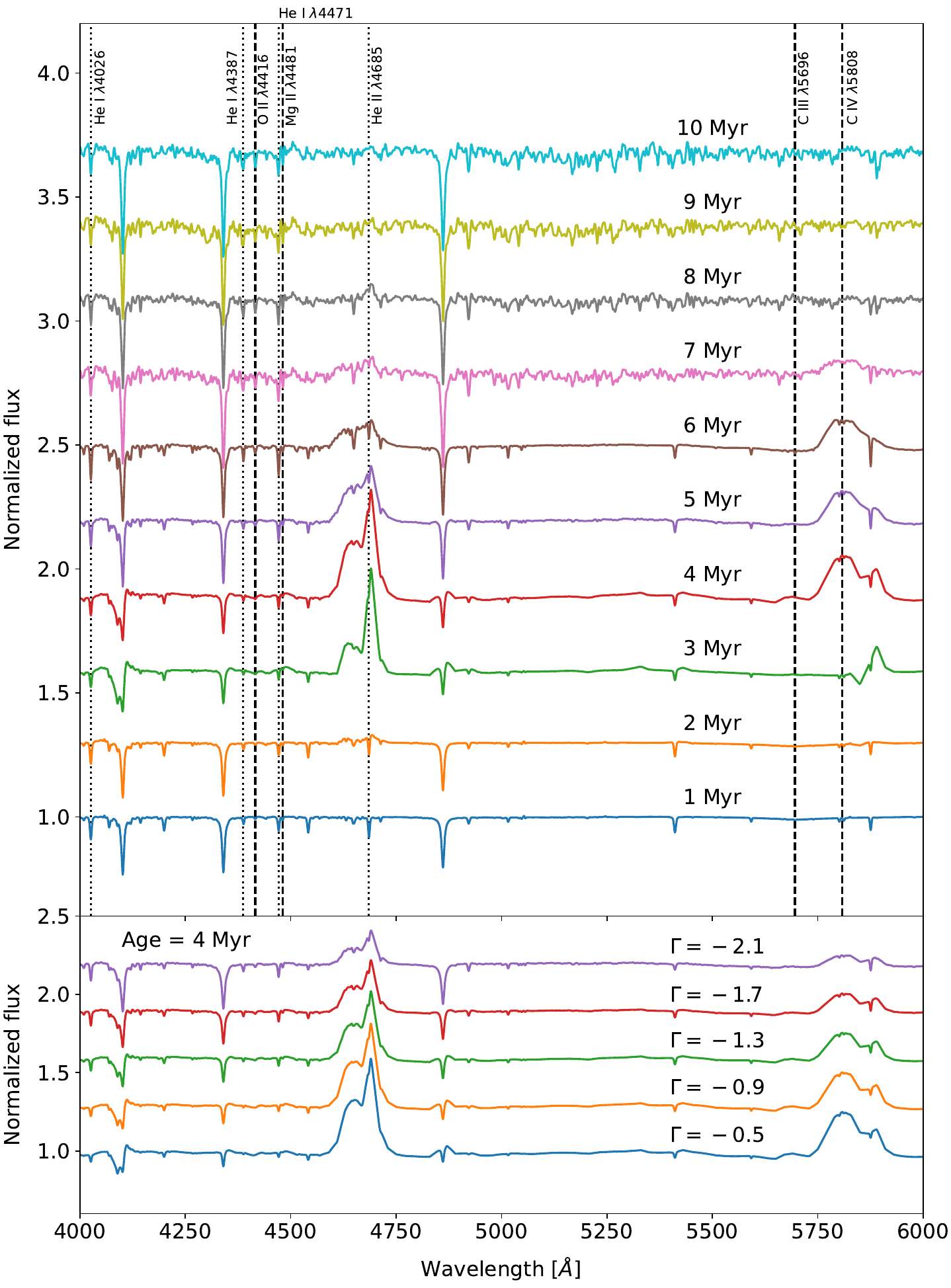}
\caption{Synthetic spectra of the model clusters 
($10^6 \ M_{\sun}$) with different ages (upper)
and underlying IMFs (lower) for the Solar metallicity. 
The adopted ages and the power-law indices of the underlying 
IMFs are shown at the top of each spectrum. In the upper 
panel, several He and metallic lines related to the ages of clusters 
are shown as the dotted and dashed lines, respectively.}\label{fig4}
\end{figure*}

\begin{figure*}[t]
\epsscale{1.0}
\plotone{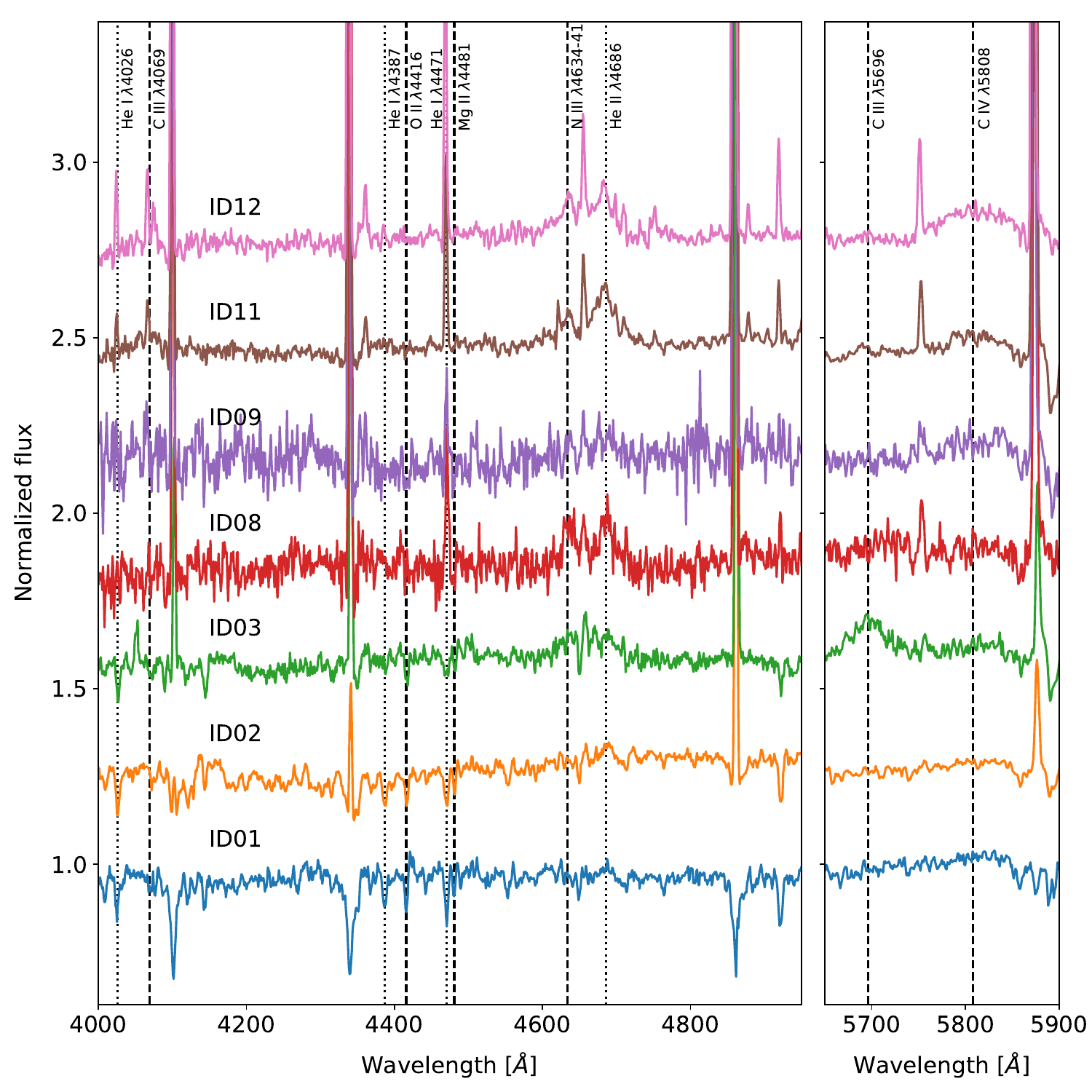}
\caption{Spectra of the observed clusters in two spectral 
windows (left and right). The IDs of 
the individual clusters are labeled in the upper left 
corner of each spectrum. The spectral lines used 
for the age estimation are displayed in the same way as 
shown in Figure~\ref{fig4}.}\label{fig5}
\end{figure*}

\subsection{Age estimation from spectral lines}\label{sec:31}
The evolutionary states of the most luminous 
stars in a given cluster determine the integrated 
spectral features. Therefore, several spectral 
lines associated with these stars can be used 
as age indicators. Figure~\ref{fig4} displays 
the synthetic spectra of the model clusters with 
a total stellar mass of $10^6 M_{\sun}$ 
generated from \texttt{STARBURST99}. We 
applied different underlying IMFs and 
ages to the individual model clusters for the Solar 
metallicity, as shown in the figure. 

The upper panel of Figure~\ref{fig4} shows 
the integrated spectra with respect to 
the ages of the model clusters for the same 
underlying IMF ($\Gamma = -1.3$). 
The most prominent features are the blue 
and red bumps at around 4686 \AA \ and 5808 \AA, respectively. 
These emission bands appear the strongest 
at 4-5 Myr and gradually weaken as the cluster ages become older, 
making them crucial for distinguishing clusters 
younger than 8 Myr. By contrast, the neutral 
He absorption lines (\ion{He}{1} $\lambda\lambda\lambda$4026, 4387, 4471) 
as well as the two metallic lines \ion{O}{2} $\lambda$4416 
and \ion{Mg}{2} $\lambda$4481 appear 
to be stronger for older clusters. 

The lower panel displays the synthetic 
spectra of the model clusters with different underlying IMFs 
at 4 Myr. The intensities of spectral lines depend 
on the underlying IMFs, but the spectral variation 
with the cluster ages is more pronounced than 
the variation of the IMFs. Thus, it is possible 
to constrain the ages of clusters based on their 
spectral features. We estimated the ages of the 
observed clusters from the criteria summarized as 
below:
\begin{enumerate}
\item $\lesssim$ 2 Myr : Weak spectral lines
\item 3 -- 4 Myr : Weak \ion{He}{1} lines, the 
absence of metallic lines, and the strongest Wolf-Rayet 
features (red and blue bumps) 
\item 5 -- 6 Myr : Metallic lines (\ion{O}{2} $\lambda$4416 and \ion{Mg}{2} $\lambda$4481) 
but weaker than \ion{He}{1} $\lambda\lambda$4387, 4471, and the weak Wolf-Rayet features 
\item $>$ 8 Myr : \ion{Mg}{2} $\lambda$4481 comparable to 
the adjacent \ion{He}{1} lines and disappearance of the blue 
and red bumps

\end{enumerate}

Figure~\ref{fig5} shows the normalized spectra of the 
observed clusters. The two clusters IDs 01 and 02 
show weak blue and red bumps, and the presence of 
\ion{He}{1} as well as metallic lines is evident. 
Therefore, they are likely older than 6 Myr but younger 
than 8 Myr. On the other hand, the spectra of the other 
five clusters (IDs 03, 08, 09, 11, and 12) are characterized 
by the blue and red bumps, indicating that these clusters are 
younger than 6 Myr. 

The spectrum of ID 03 shows three 
detectable absorption lines \ion{O}{2} $\lambda$4416, 
\ion{He}{1} $\lambda$4471, and \ion{Mg}{2} $\lambda$4481. 
\ion{C}{3} $\lambda$5696 is stronger than the blue bump, 
which implies that in this cluster, WC stars are more 
abundant than WN stars \citep{GC09}. Considering that 
the marginal timescale on which WC-type stars can be found 
is between about 6 and 7 Myr \citep{GEM12}, the age of ID 03 
is likely in this timescale.

The spectra of IDs 08, 09, 11, and 12 do not show 
detectable absorption lines of \ion{He}{1} $\lambda$4387, 
\ion{O}{2} $\lambda$4416, and \ion{Mg}{2} $\lambda$4481, 
which can be intrinsically weak. We 
introduced some noise into the synthetic spectra for 
3 and 5 Myr-old clusters, considering the SNRs of 
the observed spectra. Indeed, no absorption line was 
detected in the synthetic spectra with SNRs similar 
to the observed ones. These clusters are thus likely 
younger than 5 Myr. 

The spectra of IDs 09, 11, and 12 show a stronger red bump 
than \ion{C}{3} $\lambda$5696, which suggests 
the presence of early-type WC stars. In the 
spectra of IDs 11 and 12, the intensities of 
\ion{He}{2} $\lambda$4686 are higher than those 
of \ion{N}{3} $\lambda$4634--41 in the blue bump. 
This implies that the ages of these two clusters are about 3--4 Myr.
We estimated the age of ID 09 to be approximately 2 Myr, given 
the blue bump weaker than those of IDs 11 and 12. ID 08 
does not show the red bump, and the intensity of 
\ion{He}{2} $\lambda$4686 is comparable to that of 
\ion{N}{3} $\lambda$4634--41 in the blue bump. In addition, 
there is no detectable He and metallic absorption line. 
Therefore, ID 08 seems to be younger than 6 Myr but older than 
4 Myr.

The nebulosity around the YMCs supports our claim 
that the observed clusters are indeed young. The detection 
of several forbidden emission lines, such 
as [\ion{Fe}{3}] $\lambda$4658, [\ion{O}{3}] $\lambda\lambda$ 
4959, 5007, and [\ion{N}{2}] $\lambda$5755, and the strong 
Balmer emission lines (H$\beta$, H$\gamma$, and H$\delta$) 
implies the presence of hot gas. These emission lines are, 
in general, found in the cavities and the ridges of 
\ion{H}{2} regions \citep{OTV92,LSB18}. The origin 
of the observed emission lines is also related to \ion{H}{2} 
regions around the YMCs. The centers of the Balmer lines are 
shifted away from those of the absorption lines, which is likely 
due to the expansion of the \ion{H}{2} regions \citep{LSB18}. The YMC 
ID 01 does not show strong emission lines. We checked the public 
data cube taken with the Multi Unit Spectroscopic Explorer \footnote{\url{http://archive.eso.org/cms.html}} (MUSE). The MUSE image 
covering the H$\alpha$ emission line reveals that this YMC 
seems to be spatially associated with a giant \ion{H}{2} 
region \citep[see also the knot G in the figure 1 of][]{GMT21}. 
Since  the sky region in the GMOS slit only covers the cavity 
of the giant \ion{H}{2} region, the Balmer emission lines appears 
to be weak. We confirmed that the weak H$\beta$ emission was removed
from our spectrum after sky subtraction.

\begin{figure*}[t]
\epsscale{1.0}
\plotone{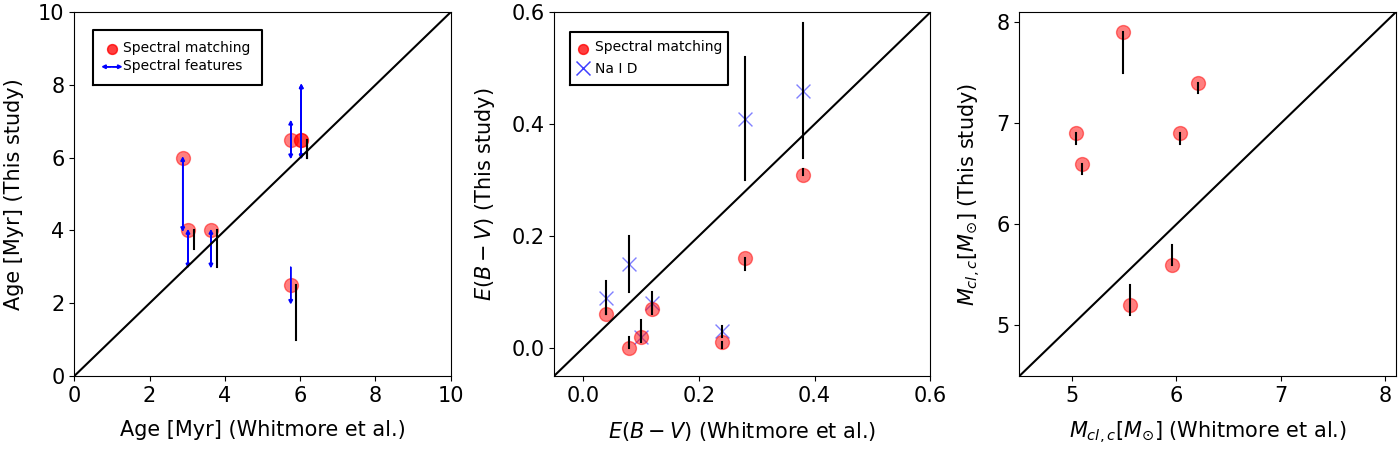}
\caption{Comparison of the parameters derived in this study 
with the results from \citet{WCS10}. 
The diagonal solid lines represent one-to-one relations for given 
parameters. The left panel compares the ages from the 
previous study with ours. The blue arrows indicate the age 
ranges estimated from the spectral features, and the red dots 
denote the ages inferred from the spectral matching. The middle 
panel shows a comparison of the reddening values. The blue crosses 
and red dots represent the results obtained from the 
\ion{Na}{1} D lines and spectral matching, respectively. The 
vertical lines represent the errors on each parameter.
The current masses of the YMCs estimated from this study 
are compared with those from the previous study 
in the right panel.}\label{fig6}
\end{figure*}

The left panel of Figure~\ref{fig6} compares the 
ages estimated by the spectral features in this study 
(blue arrows) with the estimates of \citet{WCS10}. The 
age of the clusters was found to be similar on average 
to the previous estimate, with a standard deviation of 
about 1.6 Myr.

\begin{figure*}[t]
\begin{center}
\includegraphics[width=16cm]{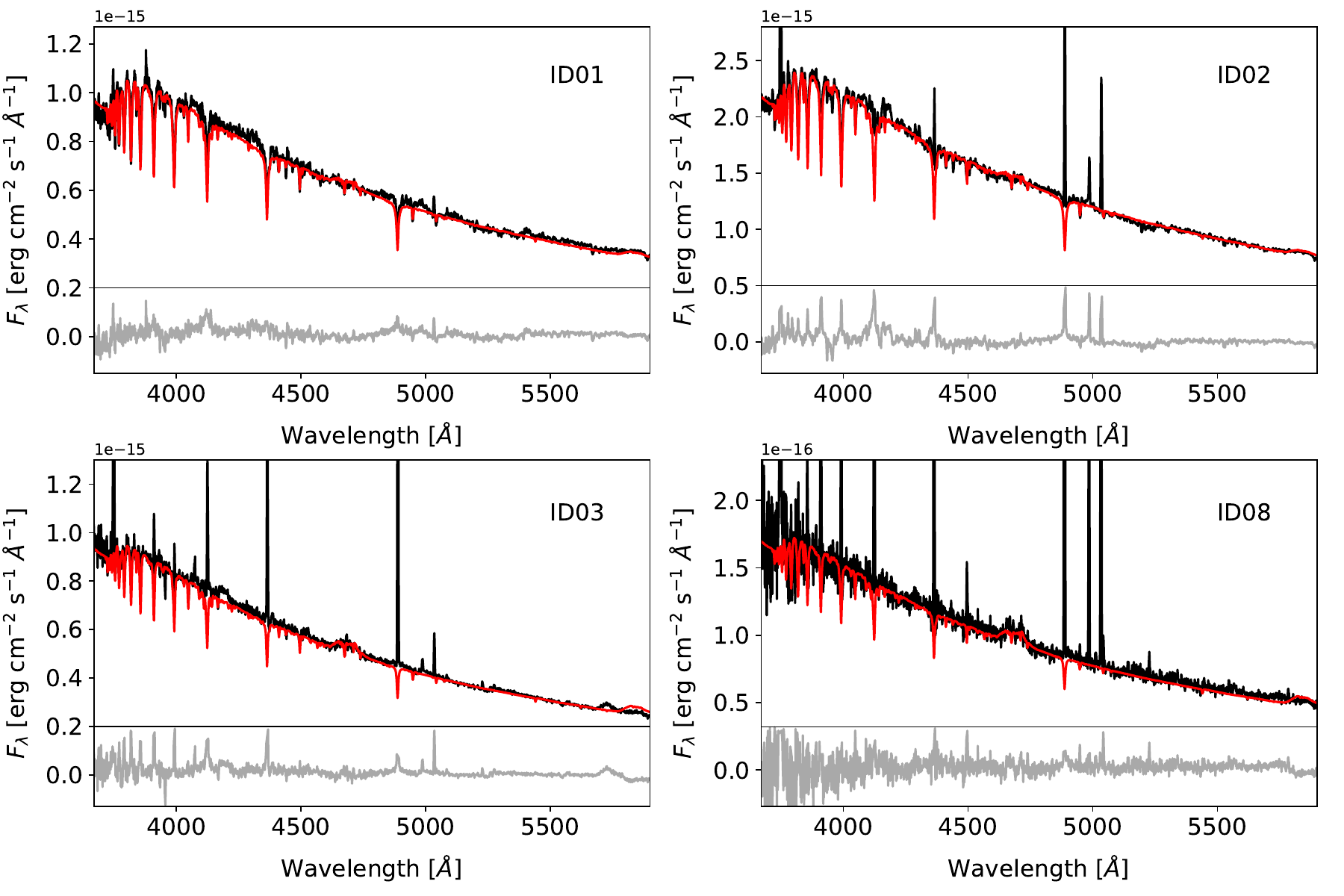}

\includegraphics[width=16cm]{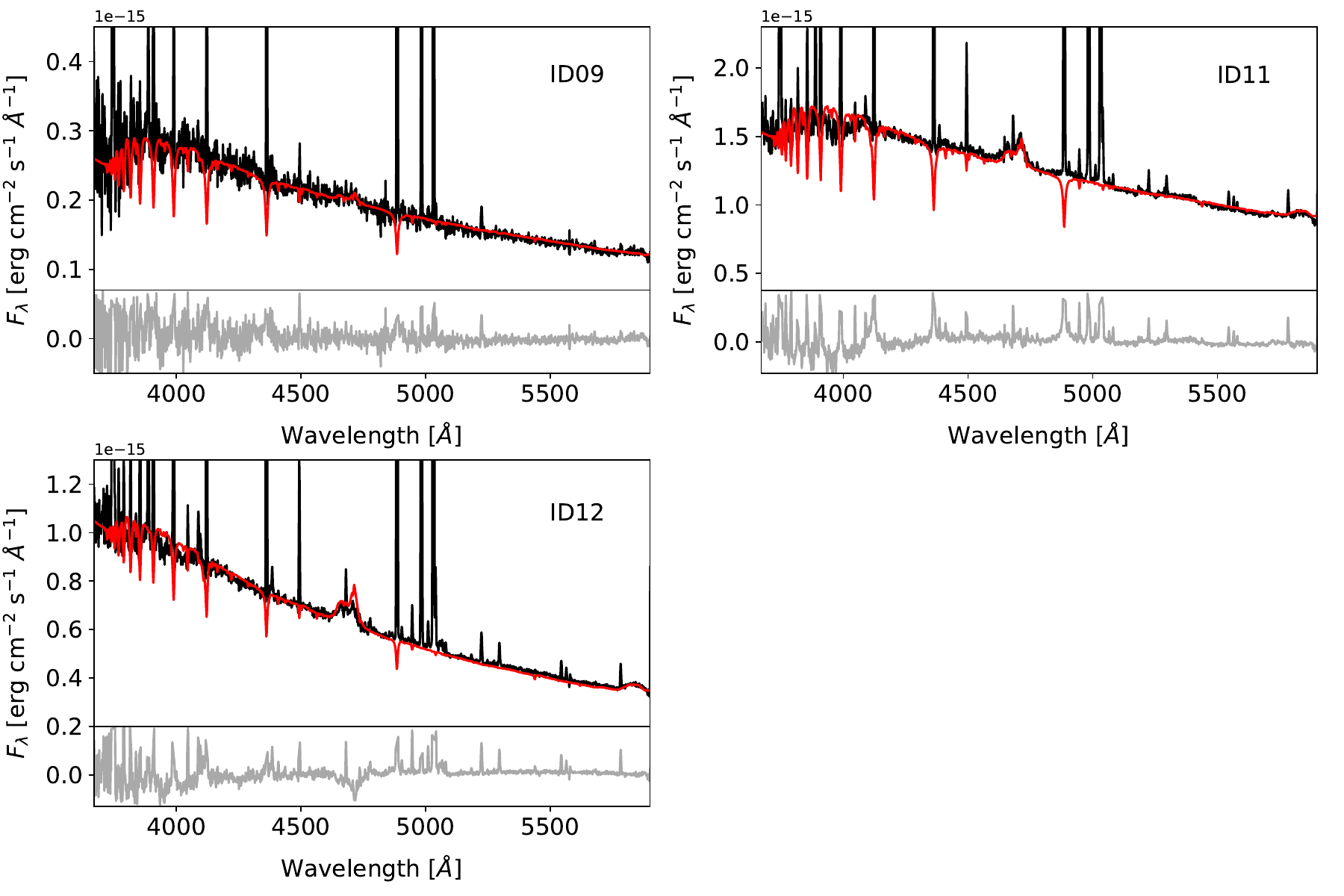}
\end{center}
\caption{Comparison of the observed spectra (black) 
with the best-fit synthetic spectra (red). The residual for 
the difference between the observed and synthetic spectra 
is shown below the horizontal line in each panel.
}\label{fig7}
\end{figure*}

\subsection{Reddening}\label{sec:32}
The light from young stellar clusters in NGC 4038/9 is 
obscured by the interstellar medium within the Galaxy
as well as their host galaxies. We adopted the 
mean reddening of $E(B-V)=0.04$ toward NGC 4038/9 from 
the Galactic reddening map \citep{SFD98,SF11}. A total 
extinction of $A_V = 0.12$ mag was adopted by assuming 
the mean total-to-selective extinction ratio of $R_V 
= 3.1$. We note that the reddening map of \citet{SF11} 
was re-calibrated by a factor of 0.86 from their previous 
reddening map \citep{SFD98}. This study adopted the more recent 
$E(B-V)$ scale of \citet{SF11}. 

The strength of the \ion{Na}{1} D lines is a useful tool to estimate 
the internal extinction of external galaxies. We adopted 
different empirical relations between extinction and the 
EWs of the \ion{Na}{1} D lines, because the line strength nonlinearly 
increases with interstellar reddening \citep{PPB12,MZM15}. 
The relations of \citet{MZM15} and \citet{PPB12} were applied 
to the less reddened cases [$E(B-V) \leq 0.08$] and more 
reddened cases [$E(B-V) > 0.08$], respectively. We identified 
the \ion{Na}{1} D lines in the observed spectra, but the spectral 
resolution of our spectra was insufficient to directly measure 
the individual line strengths. The internal reddening of the 
individual clusters was estimated by comparing the modeled 
\ion{Na}{1} D lines with the observed ones.

Based on the empirical relations of the two previous studies 
\citep{PPB12,MZM15}, we computed the EWs in the $E(B-V)$ 
range of 0.01 to 1.2 with an interval of 0.01. The \ion{Na}{1} D lines 
were modeled from the computed EWs, where the line profiles were 
assumed to be Gaussian. 
Instrumental broadening was applied to the modeled line 
profiles. These modeled lines were then subtracted from the 
observed \ion{Na}{1} D lines. We adopted $E(B-V)$, which shows the 
minimum rms value of the residuals. Since the reddening values 
of \citet{PPB12} and \citet{MZM15} were scaled to those 
of \citet{SFD98}, the $E(B-V)$ that we obtained were multiplied 
by 0.86. The propagated errors in the uncertainties of the two 
empirical relations were adopted as the errors of the reddening.

\ion{Na}{1} D lines form in the atmospheres of late-type 
stars, and therefore, these lines can be found 
in the spectra of older clusters. We investigated the 
spectra obtained from \texttt{STARBURST99} \citep{LSG99,LEM14} 
and found \ion{Na}{1} D lines for the clusters older than 7 Myr. 
Hence, the line strengths are associated with both 
reddening and the ages of clusters. Since the ages of the 
observed YMCs are younger than or close to 7 Myr, the contribution 
of the stellar absorption line to the EW of the interstellar 
\ion{Na}{1} D line may be negligible. The middle panel of 
Figure~\ref{fig6} compares the $E(B-V)$ values from this 
study measured by the \ion{Na}{1} D lines (blue crosses) 
with those of \citet{WCS10}. Our 
estimates are on average lower than theirs by 0.05. 
\citet{BTK09} also derived a reddening of 
about 0.10 toward ID 03 (T352), which is 0.05 smaller than 
our result.

\subsection{Spectral matching}\label{sec:33}
We compared the observed spectra with a grid of 
the synthetic spectra generated from \texttt{STARBURST99} 
\citep{LSG99,LEM14} to derive the physical parameters 
of the individual clusters such as age, reddening, cluster mass, 
and the underling IMF. The radial velocities of the 
individual clusters were measured using H and He lines. 
The centers of these lines were obtained from the best-fit 
Gaussian profiles, and the radial velocities of the lines
were calculated. The mean values were adopted as the radial 
velocities of the individual clusters. 

We reduced the large grid of the synthetic spectra 
to a smaller set in the expected age range of 
given clusters. The fluxes of the synthetic spectra 
were computed by dividing the luminosities of 
the spectra by $4\pi d^2$, where $d$ is the adopted 
distance of 22 Mpc \citep{SBM08}. These fluxes were 
then reddened by the internal extinction of the host 
galaxies using the Galactic extinction curve \citep{F99} 
with an $R_V$ of 3.1 \citep{GV89}, where the reddening 
estimated from the \ion{Na}{1} D lines was used for the internal 
extinction. We shifted the reddened synthetic 
spectra to the observed wavelength frame using the 
radial velocities and applied the mean Galactic reddening to 
the spectra.

For spectral matching, the $\chi^2$ values between 
the observed and synthetic spectra were computed in 
a full wavelength range. Strong H and He lines as well 
as nebular lines were masked out in the spectral 
fitting procedure. In particular, the nebular lines 
introduce slightly larger errors in the $\chi^2$ 
minimization because the synthetic spectra do 
not include such emission lines. The Balmer jump 
and the two bumps are key features in the determination 
of the underlying IMFs, and therefore, 
three spectral windows covering these features were 
weighted in the $\chi^2$ calculation. We probed the $\chi^2$ values 
adjusting the internal extinction at a given age. 
This iterative process was extended to the ranges of 
the ages estimated from the spectral features (Section~\ref{sec:31}). 

The synthetic spectra were sorted in decreasing 
order of $\chi^2$ value. We visually inspected the 
top three spectra with the minimum $\chi^2$. The synthetic 
spectra with the smallest $\chi^2$ values were 
adopted as the best-fit spectra, but in some 
cases (IDs 09 and 11) these spectra did not 
best match the two bumps. In such cases, we 
selected the synthetic spectra with the smallest 
$\chi^2$ values for the two bumps among the top three 
models. The top three models yield nearly consistent 
parameters of the observed clusters. The upper and 
lower bounds in age, reddening, cluster mass, and 
the IMF from the top three models were adopted as 
the errors of the individual parameters (vertical 
lines in Figure~\ref{fig6}). Figure~\ref{fig7} shows the spectra that best fit the observed spectra. We summarized the physical 
parameters of the YMCs in Table~\ref{tab1}.

Figure~\ref{fig6} compares the parameters of the observed 
clusters obtained by different methods. In the left panel, 
the ages obtained from the spectral matching (red dots) 
naturally overlap with the age ranges estimated from the 
spectral features (blue arrows), since the best-fit synthetic 
spectra were searched for in the expected age ranges. Therefore, 
our estimates are, on average, similar to the results of \citet{WCS10} 
for the same clusters. In the middle panel, the spectral matching 
technique (red dots) tends to yield smaller reddening 
than those obtained from the \ion{Na}{1} D lines (blue crosses), 
and also smaller than the reddening of \citet{WCS10}. 

Spectral matching yields the initial masses 
of the individual clusters because \texttt{STARBURST99} 
does not give the current masses of clusters. From the 
initial masses, we estimated the current masses ($\log M_{cl,c}$) 
using the Monte-Carlo technique. A large number of artificial 
stars were generated in a mass range from 0.1 $M_{\sun}$ to 
120 $M_{\sun}$, where the IMFs obtained from the spectral 
matching were adopted as the underlying IMFs of the individual 
clusters. The Geneva stellar evolutionary models for the 
Solar metallicity \citep{EGE12} were used to consider the 
mass loss due to stellar evolution at given ages. This 
simulation was repeated 100 times. 

As a result, the YMCs still appears to have current 
masses comparable to their initial masses in a 
logarithmic scale ($\log M_{cl,i} \approx \log M_{cl,c}$). 
The effect of stellar evolution on the cluster masses is 
less then 1\%, except for the two YMCs IDs 03 and 08. 
The right panel of Figure~\ref{fig6} compares the 
current masses of the YMCs from two studies, showing that 
70\% of the YMCs in this study have significantly higher 
masses than those from \citet{WCS10} by more than one order 
of magnitude. In the extreme case, the difference is larger 
than two orders of magnitude. We discuss the possible 
causes of the systematic differences in later section. The mass of ID 03 
($\log M_{cl,c} = 5.6$) is in good agreement with the results 
of \citet{BTK09}; however, we obtained a significantly smaller internal 
reddening [$E(B-V) \sim 0$] compared to their result and 
a shallow IMF ($\Gamma = -1.0$).

\begin{figure}[t]
\epsscale{1.0}
\plotone{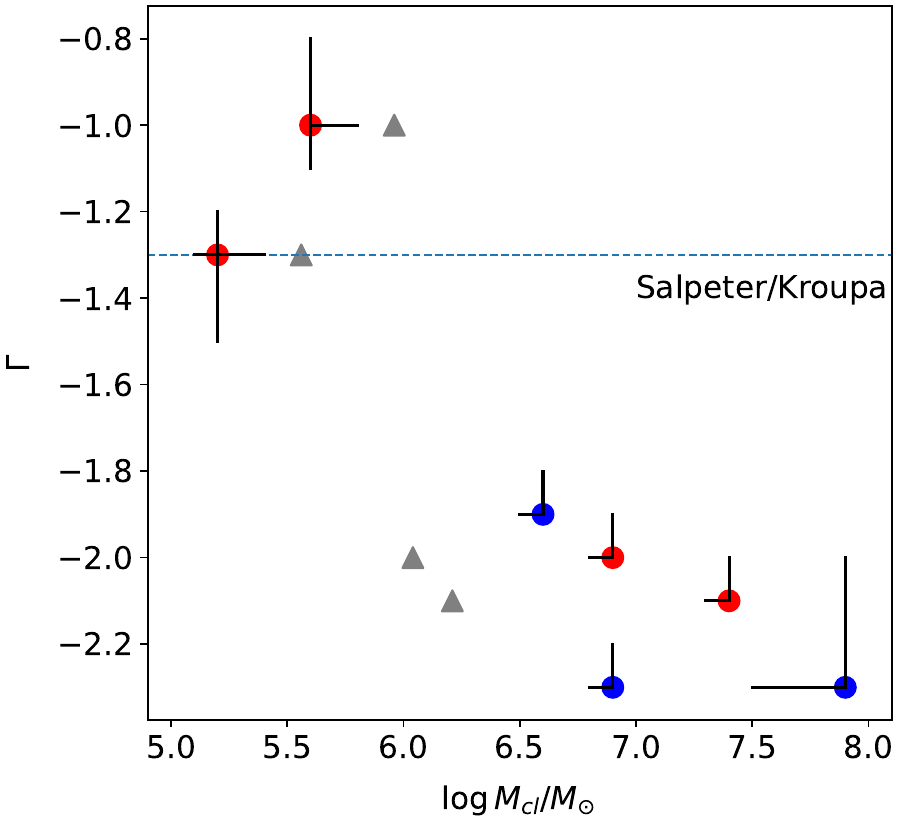}
\caption{Correlation between the cluster masses 
and $\Gamma$ of the underlying IMFs. The red and blue 
dots represent the spatially well-isolated YMCs and the 
blended YMCs with nearby sources, respectively. The 
triangle shows the relationship between the cluster masses 
from \citet{WCS10} and the underlying IMFs from this study 
for the isolated YMCs. The vertical and horizontal lines 
indicate the upper and lower bounds of cluster masses and 
$\Gamma$ obtained from the spectral matching. The horizontal 
line represents the $\Gamma$ value of $=-1.3$ for the Salpeter/Kroupa 
IMF.}\label{fig8}
\end{figure}

Most underlying IMFs inferred from the integrated 
spectra of the observed YMCs appear to deviate 
from the Salpeter/Kroupa IMF \citep{S55,K01}. 
Interestingly, we also found a correlation between 
the cluster masses and the power-law indices 
$\Gamma$, as plotted in Figure~\ref{fig8}. The masses 
of stars in more massive clusters are drawn from the 
IMF with smaller $\Gamma$, i.e. the bottom-heavy IMF. 
Given the small differences in the fundamental parameters among 
the top three models, the wide range of $\Gamma$ 
values is unlikely due to the uncertainties from the 
analysis. Therefore, our result gives an implication that 
adopting a universal form of the IMF may not always be valid. 
By contrast, this trend appears to be weaker in 
the correlation between the cluster masses estimated 
by \citet{WCS10} and $\Gamma$ obtained in this study 
(triangles).

\begin{figure*}[t]
\epsscale{1.0}
\plotone{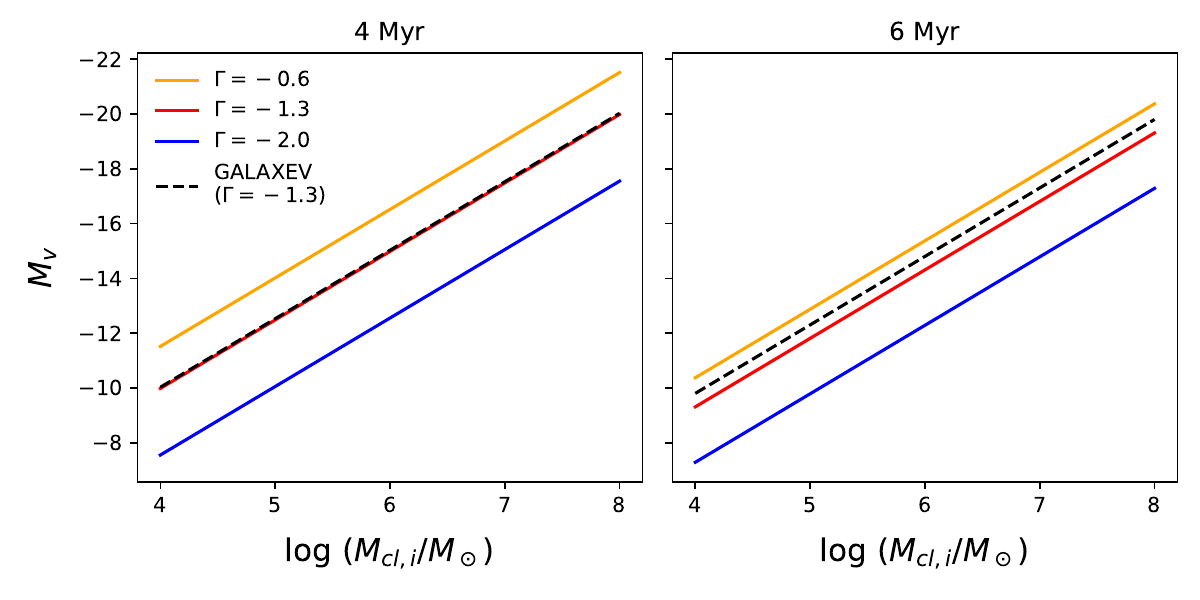}
\caption{Initial mass-luminosity relations with three different underlying 
IMFs ($\Gamma = -0.6$, $-1.3$, and $-2.0$) at 4 and 6 Myr. These relations 
were obtained from \texttt{STARBURST99} for the Solar metallicity. The adopted 
stellar evolution models consider the effects of rotation on stellar evolution 
\citep{EGE12}. The dashed line represents the initial mass-luminosity relation 
from \texttt{GALAXEV} \citep{BC03} for the Kroupa IMF.
}\label{fig9}
\end{figure*}

\begin{deluxetable*}{lclcccccccc}
\tabletypesize{\footnotesize}
\tablewidth{0pt}
\tablecaption{Physical properties of seven YMCs in NGC4038/9 \label{tab1}}
\tablehead{
\colhead{ID} & \colhead{$V$} & \colhead{$E(B-V)_{\mathrm{Na ~I}}$} & \colhead{$E(B-V)_{\mathrm{mat}}$} & \colhead{Age$_{\mathrm{spec}}$ [Myr]} & \colhead{Age$_{\mathrm{mat}}$ [Myr]} & \colhead{$\log M_{cl,i}$} & \colhead{$\log M_{cl,c}$} & \colhead{$\Gamma$} & \colhead{SNR} 
 & \colhead{Whitmore ID}  }
\startdata
01 & 17.57 & $0.09$ $(0.03)$    & $0.06^{+0.01}_{-0.00}$ & 6--8 & $6.5^{+0.0}_{-0.5}$ & $6.9^{+0.0}_{-0.1}$ & $6.9$ & $-2.0^{+0.1}_{-0.0}$ & 40 & 50776\\
02 & 16.66 & $0.08$ $(0.02)$    & $0.07^{+0.01}_{-0.00}$ & 6--8 & $6.5^{+0.0}_{-0.5}$ & $7.4^{+0.0}_{-0.1}$ & $7.4$ & $-2.1^{+0.1}_{-0.0}$ & 80 & 36094\\
03 & 17.83 & $0.15$ $(0.05)$    & $0.00^{+0.02}_{-0.00}$ & 6--7 & $6.5$               & $5.7^{+0.1}_{-0.1}$ & $5.6$ & $-1.0^{+0.2}_{-0.1}$ & 80 & 38220(38377, 38305, and 38195)\\
08 & 19.55 & $0.03$ $(<0.01)$   & $0.01^{+0.00}_{-0.01}$ & 4--6 & $6.0$               & $5.2^{+0.2}_{-0.1}$ & $5.2$ & $-1.3^{+0.1}_{-0.2}$ & 25 & 55784\\
09 & 18.62 & $0.41$ $(0.11)$    & $0.16^{+0.00}_{-0.02}$ & 2--3 & $2.5^{+0.0}_{-1.5}$ & $6.9^{+0.0}_{-0.1}$ & $6.9$ & $-2.3^{+0.1}_{-0.0}$ & 45 & 22079(22109)\\
11 & 16.34 & $0.46$ $(0.12)$    & $0.31^{+0.01}_{-0.00}$ & 3--4 & $4.0^{+0.0}_{-0.5}$ & $7.9^{+0.0}_{-0.4}$ & $7.9$ & $-2.3^{+0.3}_{-0.0}$ & 60 & 15502(15492 and 15466)\\
12 & 17.22 & $0.02$ $(<0.01)$   & $0.02^{+0.03}_{-0.00}$ & 3--4 & $4.0^{+0.0}_{-1.0}$ & $6.6^{+0.0}_{-0.1}$ & $6.6$ & $-1.9^{+0.1}_{-0.0}$ & 90 & 19459(19416, 19374, 19435, and 19394) 
\\
\enddata
\tablecomments{Column (1) : Cluster ID. Column (2) : $V$ magnitude from synthetic photometry.
Columns (3) and (4) : Reddening estimated from the \ion{Na}{1} D line and the spectral matching. 
Columns (5) and (6) : Ages estimated from the spectral features and the spectral matching. 
Columns (7) and (8) : The initial cluster mass and current cluster mass in a logarithmic scale. 
Column (9) : The power law index $\Gamma$ of the IMF. Column (10) : Signal-to-Noise Ratio. 
Column (11) : The IDs from \citet{WCS10}. The numbers in parentheses represent the IDs of neighboring sources.}
\end{deluxetable*}

\section{Discussion}\label{sec:4}
The fundamental parameters of seven YMCs in NGC 4038/9 
were derived by comparing their integrated spectra with 
those synthesized from a simple stellar population model. 
As a result, our results suggest a variation of the stellar 
IMFs. Here, we discuss the several sources that can 
cause uncertainties in our analysis. The first source 
is the SNR of a given spectrum. A low SNR can blur spectral 
features, increasing the uncertainty in derived physical 
parameters. 
To investigate the effects, 
we introduced random errors corresponding to final SNRs 
into the synthetic spectra that best matched 
the observed spectra, and then retrieved the physical 
parameters of YMCs from the noise-added spectra. This simulation 
was repeated 100 times. As a result, small errors (0.1 dex 
in $\log M_{cl,i}$ and 0.1 in $\Gamma$) occurred only for 
ID 08 with SNR of 25. 
Therefore, the SNR may not introduce large errors on the physical 
parameters of YMCs. 

The second source is 
reddening correction. The internal reddening 
obtained from the \ion{Na}{1} D line tends to be larger than 
that from spectral matching. 
One possibility for this difference is that the strength 
of the \ion{Na}{1} D line is not a precise 
reddening tracer. Indeed, there is a weak 
correlation between the \ion{Na}{1} and \ion{H}{1} 
column densities due to the large scatter in the 
derived column densities \citep{MZM15}. 
The diffuse interstellar band (DIB) at 5780 \AA\ 
traces the reddening fairly well \citep{H93,FYM11,
LMZ15}, and there is a good correlation between 
the strength of the DIB $\lambda 5780$ and the 
\ion{Na}{1} column density. However, the 
relationship between the strength of the 
DIB and reddening as well as the \ion{H}{1} 
column density does not appear to hold for strong 
ultraviolet radiation fields \citep{FYM11,W14}. 
If the behavior of \ion{Na}{1} is similar to that 
of the DIB $\lambda 5780$, the \ion{Na}{1} D line 
may not precisely trace reddening toward the 
YMCs containing a number of ionizing sources.

The spectra of the YMCs are reddened with $E(B-V)$ 
ranging from 0.02 to 0.46 in addition to the Galactic 
reddening. The correction for the internal reddening is 
complicated. The light from the individual YMCs is obscured 
by the diffuse interstellar medium in their 
host galaxies as well as the remaining natal clouds
(or intracluster medium). Although the contribution 
of the diffuse interstellar medium in NGC 4038/9 
to the internal reddening may not be high as these 
host galaxies have a face-on morphology, the natal 
clouds still contain a large fraction of dust grains. Furthermore, 
the inhomogeneous distribution of dust grains across the remaining 
clouds results in differential reddening. It is practically 
impossible to correct for differential reddening in the 
integrated spectra of unresolved stellar systems. Therefore, 
some of the differences between the observed and 
best-matched spectra may be related to differential 
reddening.

The Galactic mean $R_V$ was adopted for calculating the internal 
extinction $A_V$. This $R_V$ can also result in systematic uncertainties 
in our analysis. A number of photometric studies have 
confirmed that the abnormal reddening law deviates from the Galactic 
mean $R_V$ for very young active star-forming regions in 
the Galaxy, e.g., the Orion Nebula Cluster \citep{J67,
CCM89,DTC16}, Carina Nebula \citep{HLC23}, Westerlund 2 
\citep{HPS15}, NGC 281 \citep{KLB21}, NGC 6530 \citep{FGH12}, 
and NGC 6611 \citep{CW90}.

However, there is a tendency that open clusters older 
than 3 -- 5 Myr do not show the abnormal reddening 
law, e.g., IC 1848 \citep{LSK14}, IC 1805 \citep{GV89,SBC17}, 
and NGC 6231 \citep{SSB13}. In particualr, the $R_V$ in the Carina 
Nebula shows a clear spatial variation with respect to 
the age of young clusters \citep{HLC23}. Anomalous 
extinction is found for young open clusters Trumpler 14, 
16, and Collinder 232 in the southern Carina Nebula, 
but $R_V$ becomes normal for the older cluster Trumpler 
15 in the northern region. These facts imply that 
the timescale for the evolution of dust grains in size 
is approximately 3 -- 5 Myr, and it may also depend on 
the amount of ultraviolet radiation flux and the 
metallicity of star-forming regions. 

With the exception of ID 09, the observed YMCs are 
older than 4 Myr, and therefore, the $R_V$ toward these 
clusters is thought to be normal. Even if anomalous extinction 
affects the integrated light from the YMCs, the effect 
on our results would not be severe. The increase in $R_V$ 
due to dust growth mitigates the effect of 
wavelength-dependent selective extinction in optical 
passbands \citep{F99}. For a quantitative test, we applied 
an $R_V$ of 5 to the spectra of ID 11 with the largest $E(B-V)$ and 
searched for the best-matched synthetic spectra. This 
resulted in a small change in the fundamental parameters 
of the YMC: an age of 0.5 Myr, a logarithmic cluster 
mass of 0.1 dex, and a power-law index of 0.3.

The third source is the adopted metallicity. 
The synthetic spectra generated from a simple stellar 
population model for the Solar metallicity were matched  
to the observed spectra. We found no evidence that the 
YMCs have different metallicity from the Solar 
value. Synthetic spectra for sub-Solar metallicity 
($Z = 0.002$) were compared with the observed spectra, 
but none of the synthetic spectra fit the Wolf-Rayet 
bumps. This result is consistent with the observation 
that the Wolf-Rayet wind is weaker in low-metallicity 
environments \citep{CH06}. Chemical analyses from 
previous studies also support our claim. \citet{BTK09} 
analyzed the integrated spectra of several stellar 
clusters in NGC 4038/9 and found that these clusters 
have the Solar metallicity. \citet{GMT21} also confirmed 
the Solar metallicity from the oxygen abundance of 
Wolf-Rayet stars in the same galaxies. 

The fourth source is the degeneracies among 
the physical parameters of YMCs. Since the integrated 
spectral features depend 
on the cluster age and the underlying IMF, as 
shown in Figure~\ref{fig3}, our results can still 
be affected by the degeneracies between the 
two parameters. There are possibly multiple 
models explaining the observed spectra. We tested 
the possibility that the stellar population in 
the observed YMCs is drawn from the Salpeter/Kroupa IMF. 
A simple manner to do this is to apply the 
fundamental parameters derived by \citet{WCS10} 
to the synthetic spectra and compare them with 
the observed spectra, because they analyzed the 
photometric data by means of a simple stellar 
population model adopting the Salpeter/Kroupa IMF. 
However, there is no synthetic spectrum that 
matches the observed spectra in both stellar 
evolution models with and without rotation.

For the five YMCs with the bottom-heavy 
IMFs, we adopted the Salpeter/Kroupa 
IMF as the underlying IMF and examined the 
best-matched synthetic spectra in the ranges 
of ages and reddening values. However, the 
newly matched synthetic spectra do not fit 
the Balmer jumps or the overall slope of 
the observed spectra any better than the 
synthetic spectra with the bottom-heavy IMFs. 
Therefore, our results may not originate 
from the degeneracies between the age and 
the underlying IMF.

The fifth source is the blending effect 
with neighboring objects. The masses of five YMCs 
are significantly larger than those estimated 
by \citet{WCS10}. Since these YMCs occupy 
the high-mass regime in the cluster mass-IMF 
correlation (Figure~\ref{fig8}), the reliability 
of the derived cluster masses should be scrutinized. 
We confirmed that three YMCs (IDs 09, 11, and 12) have at least 
one bright neighboring source in the high-resolution HST 
images. Since these sources were not resolved in our 
seeing-limited pre-image, the integrated fluxes of 
these YMCs are likely overestimated compared to those 
from \citet{WCS10}, leading to the high cluster masses. 
In fact, the $V$ magnitudes of these three YMCs obtained 
from the synthetic photometry are smaller (brighter) than 
those of the previous study. If the individual 
YMCs and their neighboring sources are part of 
stellar associations or complexes, it is possible 
that we have inferred the masses and the underlying IMFs 
of such large-scale stellar systems. 

On the other hand, the other two YMCs (IDs 01 and 02) 
are spatially well-isolated without bright neighboring 
sources. In addition, their brightness in the $V$ band 
are comparable to the previous measurements. We examined 
the systematic error originating from the adopted stellar 
population models as the sixth source. Figure~\ref{fig9} shows a comparison of 
the mass-luminosity relations from \texttt{STARBURST99} 
(red solid line -- \citealt{LSG99,LEM14}) and \texttt{GALAXEV} 
(black dashed line -- \citealt{BC03}) for the Kroupa IMF. At 4 Myr, 
the two relations are almost the same, while the cluster 
masses estimated from \texttt{GALAXEV} are 0.2 dex 
smaller than those from \texttt{STARBURST99} for a given 
luminosity at 6 Myr. There are systematic differences 
between the mass-luminosity relations from the 
two models, but the systematic errors associated with 
the adopted models are insufficient to explain the significant 
mass differences between the previous study and ours 
for the two clusters.

We suggest that the differences in cluster masses may be 
attributed to the different underlying IMFs. Figure~\ref{fig9} 
displays the mass-luminosity relations for three different 
underlying IMFs (orange, red, and blue solid lines) 
at 4 and 6 Myrs. Younger clusters tend 
to be more luminous than their older counterparts 
for the same masses. At the same age, the clusters with 
the bottom-heavy IMFs are less luminous than the clusters 
with the shallower IMFs. Therefore, there are several 
possibilities for inferring 
cluster masses from a given luminosity. 

\citet{WCS10} estimated the physical parameters of clusters 
by comparing the integrated broadband photometric data with those 
predicted from the adopted stellar population model. 
Here, we emphasize that the broadband photometry is 
insensitive to the spectral features related 
to the IMFs. There are, thus, degeneracies among 
cluster ages, underlying IMFs, and cluster masses. We 
have shown that cluster ages and underlying IMFs can be 
constrained by probing the variation of spectral features 
and comparing them with the synthetic ones. Spectral analysis 
can provide a better estimate of the cluster mass. 

The YMC ID 02 is the most massive star cluster ($2.5 \times 10^7 
\ M_{\sun}$) among the four isolated YMCs. The large amount of 
line broadening due to the rapid rotation of massive stars 
hinders the estimation of the dynamical mass of this cluster. The 
assumption of this dynamical mass estimation is that a given 
cluster is in a virial state, which does not hold for 
very young clusters. We surveyed the literature to determine 
whether our estimate is acceptable within the upper bound of 
the cluster mass. The upper limit of star clusters is known to 
be about $\lesssim 10^8 \ M_{\sun}$ \citep{NVK19}. Indeed, the 
dynamical mass of W3 in NGC 7252 were estimated to be 
$8.0\times10^7 \ M_{\sun}$ \citep{MBS04}. \citet{BSG06} reported 
that the total masses of two individual clusters G114 in NGC 1316 and W30 
in NGC 7252 exceed $10^7 \ M_{\sun}$. Given the old ages 
of these clusters, their initial masses may be about 
$10^8 \ M_{\sun}$. \citet{RBD15} performed hydrodynamic simulations 
of star cluster formation in merging galaxies such as NGC 4038/9 
and found significantly massive clusters with total stellar masses up to 
$\sim 10^8 \ M_{\sun}$. The results of these previous studies 
support that the mass of the YMC ID 02 is acceptable.

We have discussed the impacts of several sources 
and assumptions such as reddening, the adopted metallicity, stellar 
population models, and the degeneracy among the fundamental parameters on 
our results. There are practical limitations in the 
proper correction for the differential reddening; 
however, it is difficult to quantitatively infer the 
uncertainty due to this effect. On the other hand, 
the other sources and assumptions do not severely 
change our results. We caution about the blend 
of multiple sources in the seeing-limited images.  
The YMC IDs 09, 11, and 12 in our sample are subject to 
blending of neighboring sources. Nevertheless, the correlation 
between the masses and the underlying IMFs of the other 
YMCs still appears to hold (Figure~\ref{fig8}). 

\section{Summary}\label{sec:5}
We investigated the integrated spectra of seven YMCs 
in NGC 4038/9 to infer their underlying 
IMFs. The features of the observed spectra were 
interpreted by means of the simple stellar population model 
\texttt{STARBURST99} \citep{LSG99,LEM14}. The fundamental 
parameters of the YMCs were inferred from the synthetic 
spectra that best matched the observed spectra. 

The most prominent spectral features associated 
with stellar content of the YMCs 
are the blue and red bumps originating from Wolf-Rayet 
star populations. The presence of these features indicates 
that the YMCs are younger than 8 Myr. The ages of the 
YMCs estimated from the best-matched synthetic spectra 
range from 2.5 Myr to 6.5 Myr, which is on average similar to the 
results of \citet{WCS10}.

The internal reddening $E(B-V)$ obtained from 
the \ion{Na}{1} D lines ranges from 0.02 to 0.46. We 
adopted the reddening values as the initial values 
for spectral matching and redetermined the 
reddening that matches the slopes of the observed 
spectra. The reddening values obtained in \citet{WCS10} 
lie between those determined by the \ion{Na}{1} D lines 
and the spectral matching.

The masses of the observed YMCs are higher than 
$10^5 \ M_{\sun}$ and smaller than $10^8 \ M_{\sun}$. 
Hence, it is confirmed that they are, indeed, massive 
stellar clusters. The YMCs appear to have various 
underlying IMFs. The power-law index $\Gamma$ ranges 
from $-1.0$ to $-2.3$, which is equivalent to 2.0 
to 3.3 in $\alpha$. More massive clusters tend to 
have bottom-heavy IMFs. Considering these results, 
the implication is that the choice of the 
underlying IMF is crucial in analyzing 
unresolved stellar populations. However, 
several of our most massive clusters are contaminated 
and might have their masses overestimated, and the 
sample size is insufficient. A systematic survey 
of young stellar clusters 
in several external galaxies with more diverse environments 
will further clarify our results.

\begin{acknowledgments}
The authors thank the anonymous referee for 
constructive comments and suggestions. The authors 
would also like to express thanks to Dr. Sang-Hyun Chun 
and Professor Dohyeong Kim for their help. This paper has made 
use of data obtained under the K-GMT Science Program (PIDs: GEMINI-KR-2022A-004 
and GEMINI-KR-2023A-004) supported by the Korea Astronomy and Space Science
Institute (KASI) grant funded by the Korean government (MSIT;
No. 2023-1-860-02, International Optical Observatory Project). 
This research has also made use of the SIMBAD database,
operated at CDS, Strasbourg, France. This work was supported by 
the National Research Foundation of Korea (NRF) grant funded by the Korean 
government (MSIT; grant No. 2022R1C1C2004102). 
This research of H.-J. K. was supported by Basic 
Science Research Program through the National Research 
Foundation of Korea (NRF) funded by the Ministry of 
Education (RS-2023-00246733).
\end{acknowledgments}

\vspace{5mm}
\facilities{Gemini South:8.1m}


\software{{\tt Astropy} \citep{astropy:2013,astropy:2018,astropy:2022}, {\tt NumPy} \citep{HMvdW20}, {\tt SciPy} \citep{VGO20}, {\tt Pyphot} \citep{F22}}








\end{document}